# The removal of the polarization errors in low frequency dielectric spectroscopy

Camelia Prodan[1] and Corina Bot[1]

Physics Department, New Jersey Institute of Technology, Newark, NJ 07102

Electrode polarization error is the biggest problem when measuring the low frequency dielectric properties of electrolytes or suspensions of particles, including cells, in electrolytes. We present a simple and robust method to remove the polarization error, which we demonstrate to work on weak and strong ionic electrolytes as well as on cell suspensions. The method assumes no particular behavior of the electrode polarization impedance; it makes use of the fact that the effect dies out with frequency. The method allows for direct measurement of the polarization impedance, whose behavior with the applied voltages, electrode distance and ionic concentration is investigated.

Low frequency impedance spectroscopy (or dielectric spectroscopy) is an experimental technique that records the low frequency variation of the impedance (or the complex dielectric permittivity) of a sample as function of frequency. Because is noninvasive, this method is widely used in many areas among which biophysics, pharmacology and geophysics. For example impedance spectroscopy have been used to monitor the living cell and plasma membrane properties or protein activity.[1-4] Impedance measurements were used to characterize the wound healing process of monolayer cell tissues.[5] The single cell DNA content was reported to be determined using impedance monitoring.[6] In Pharmacology, cellular dielectric spectroscopy has been used as a label free technology for drug discovery.[1] In geophysics low frequency impedance measurements are used to look at mineral pore size.[7,8] However, the experimental research is mostly qualitative (comparison between impedances as different processes take place), despite the fact that there are theoretical studies,[9,10] that could be used for quantitative analysis. The quantitative interpretation of the measurements is prevented by contamination of the signal with the electrode polarization. To obtain the true dispersion curves and thus to physically interpret the measurements, the polarization error has to be removed from the measurements.

The available methods to remove the polarization effect work in special conditions and/or at narrow frequency ranges.[11-16] Most of them assume a specific behavior of the polarization impedance. A detailed presentation of the existing electrode polarization removal methods can be found in Refs.[17-19].

The advantages of the technique presented in this paper are that (i) is easy to apply, (ii) no need for calibrations at any step and (iii) no assumption on the behavior of the electrode polarization impedance. This allows for applications in all conditions and frequency ranges from zero to gigahertz.

The following three paragraphs present a description of the experimental set up and its testing.

a) *Description of the experimental set up.* We use the same experimental setup as in Ref.[20]. The solution to be measured is placed between two parallel gold-plated electrodes that are enclosed in a cylindrical glass tube. The distance between the capacitor's plates is controlled with a micrometer. The results reported here were obtained with electrodes of 28.1 mm in radius. The distance between the electrode plates was varied from 1 mm to 10 mm. We measure the complex transfer function is equal to: $T = R_2/Z$, where Z is the complex impedance of our measuring cell and $R_2$ the reference resistor. $R_2$ = 100 Ω in these experiments. If the impedance $Z$ were not contaminated by the polarization effects and other possible factors, then the complex dielectric function, $\varepsilon^* = \varepsilon + \sigma/j\omega$, of the solution could be calculated from:

$$Z = \frac{d/A}{j\omega\varepsilon^*} \text{ or } \varepsilon^* = \frac{d/A}{j\omega R_2}T. \quad (1)$$

b) *Simple tests.* Before discussing the signal contamination by the polarization effect, it is instructive to see how the setup performs without any post-data processing. The measurements of individual or combined electronic components such as resistors and capacitors proved that there is very little noise in the data and the measured values are within 1% of the nominal values, even at the very low frequency range. Measurements of the value of the relative permittivity of air, as extracted from the measured impedance Z with measuring cell filled with air, gave values remarkably closed to 1, given that no calibration was used in this experiment.

c) *Samples.* The measurements reported here are for Millipore water, saline water with lower and

then higher concentrations of KCl, buffer solutions and live cells suspended in water.

The following three paragraphs describe the methodology for measuring $Z_p$ and removing it from the raw data.

a) *Polarization effect.* Although the setup is capable of highly accurate impedance measurements at low frequencies, the measured impedance Z doesn't necessarily reflect the true impedance of the solution. It is now well known that, especially at low frequencies, Z will be contaminated by the so-called "polarization" impedance, $Z_P$. This effect is due to ionic charges accumulated at the interface between the fluid and metallic electrodes and was carefully studied in several works. Here we mention that, because the effect takes place at the interfaces, the polarization impedance appears in series with the true impedance of the solution, $Z_s$. In other words, the measured impedance is the sum of the two: $Z=Z_s+Z_p$. Thus, if one develops a method to measure $Z_p$, the intrinsic impedance of the solution can be obtained by subtracting $Z_p$ from the measured Z.

b) *Measuring Zp.* We will use Millipore water to exemplify the method. The dielectric function of pure water is known to be constant ($\varepsilon_r$=78) for a frequency range spanning from 0Hz to several GHz. However, if we compute the dielectric function using Eq. 1, we obtain the graph shown in Figure 2 k (red line). The graph shows that from approximately 1kHz and up, $\varepsilon_r$ agrees extremely well with the nominal value of 78, while from 1kHz down, $\varepsilon_r$ starts increasing quite strongly. The anomalous behavior of $\varepsilon_r$ at these low frequencies is quite typical and it is precisely due to the polarization effects. But the main observation is that above 1kHz the effect is practically gone. If we plot the real and imaginary parts of the impedance Z (red lines in Figure 2 a and f), and the ideal impedance of the Millipore water, computed as: $Z_s = d/j\omega\varepsilon^* A$, where $\varepsilon^*$ is the nominal complex dielectric function (blue lines in Figure 2 a and f), we see that the two graphs match quite well above 1kHz. In fact, if we fit the experimental values of Z, above 1kHz, with a frequency dependent function

$$Z_{fit}(\omega) = \frac{d/A}{j\omega\varepsilon + \sigma}, \quad (2)$$

with ε and σ as fitting parameters, we obtain ε=78 and σ=0.000070 S/m, in very good agreement with the nominal values. But since it is known that the dielectric function and conductivity of ionic solutions are constant all the way to zero frequency, we can extrapolate $Z_s$ to lower frequencies. The difference between the measured $Z$ and $Z_s$ is precisely the polarization impedance $Z_p$. A plot of $Z_p$ is show in the inset of Figure 2 f (green line).

*The following methodology emerges:* To measure the polarization impedance for an unknown ionic solution one can do the following:

i) Fit the experimental data for $Z$ with the function $Z_{fit}$ given in equation (2), by giving a large weight to the high frequency data and a low, or almost zero weight to the low frequency data.

ii) From the fit, determine the true dielectric constant ε and conductivity σ of the ionic solution.

iii) Extrapolate $Z_s$ all the way to 0 Hz.

iv) Compute $Z_p$ as the difference between $Z$ and the extrapolated $Z_s$.

c) *Behavior of Zp.* We will use the mili-Q water to answer several important questions about the polarization impedance. $Z_p$ depends, in general, on the distance $d$ between the capacitor plates, on the applied voltage and on frequency. Figure 1 maps the dependence of Zp for mili-Q water on all these tree parameters. The methodology outlined above was applied for three values of the distance between capacitor plates, $d$=1, 3, 5mm. For each $d$, the experiments were repeated with five values of applied voltages per c[21]entimeter: 0.1, 0.3, 0.5, 0.7 and 0.9V/cm. The applied electric fields were kept below the linear regime limit of 1V/cm. The steps i)-iii) were applied to resulting 30 measurements. Following this procedure, it we found the following.

In all our measurements, we find that $Z_P$ is mainly reactive, in line with several other works[22]. Then, in the log-log plot of the insets in Figure 2 the curves appear linear for a wide range of frequencies, implying that $Z_P$ goes with the frequency as a power law. This is again in line with previous works.[23] The power law can be easily computed from the graphs. It is notable that, this very specific behavior of $Z_p$ was used in the past to remove the polarization impedance.[11, 17] We can confirm now that both, the amplitude and the exponent of the power law are weakly dependent on the applied voltage.

Another important finding is that $Z_P$ saturates at large values of $d$. This fact justifies the use of an existing method called "electrode distance variation technique" to correct for the polarization effect.[24] We found that this simple technique can be applied for electrolytes with low conductivity; for higher conductivity, the method isn't reliable.

In the following three paragraphs we demonstrate that the methodology is robust and that it can be applied to various ionic solutions. In particular, we describe how to remove the polarization effect for live cell suspensions.

a) *Weak ionic solutions.* In Figure 2 we present an application of the methodology to electrolytes with low conductivity, more precisely, water with low concentrations of KCl. From left to right, the different columns in Figure 2 refer to 0 uM KCl, 1um KCl, 5 uM KCl, 10 um KCl and 20 uM KCl. It is well known that, when increasing the KCl concentration, the dielectric function of the solution should remain constant at 78, while large increases in the conductivity should be observed. In all columns, we can see an almost perfect match between the real parts of $Z$ and $Z_{\text{fit}}$. The fit is also perfect for the imaginary parts of $Z$ and $Z_{\text{fit}}$, if we look above 1kHz. The fitting provided the following values: $\varepsilon_r = 78 \pm 1$ and $\sigma = 0.00011$, 0.00020, 0.00034 and 0.00056 S/m for the five cases, respectively. To these significant digits, same values of $\sigma$ have been measured using a regular conductivity probe. This demonstrates that, by an automated implementation of the steps i)-iv) described above, we were able to remove the polarization effects and obtain the true dielectric function and conductivity of all these solutions.

b) *Strong ionic solutions.* Low frequency dielectric spectroscopy has applications in many fields, but we are primarily interested in biological applications, such as measuring the dielectric functions of live cell suspensions. Now, many physiological buffers are known to have huge conductivities. Thus, in order to apply our method to live cells, we must demonstrate that it works for strong ionic solutions.

As the conductivity of the electrolyte increases so does the frequency limit where the polarization effect highly contaminates the data. Since our method relies on the information contained in the frequency domain where the polarization effects are small, the experimental data must contain a good part of this domain. Thus, for strong ionic solutions, we had to go to much higher frequency. This why, in these experiments we replaced the initial signal analyzer (SR 795) with another one (Solartron 1260), which allowed us to record data in the frequency window from 0Hz to $10^6$Hz.

We give an application of the methodology to water and KCl, at much higher concentrations than before, namely 0.1, 0.5 mM, and to HEPES with 5 mM concentration. The dielectric functions before and after polarization removal are shown in Figure 3 left. For KCl solutions, the methodology provides the following values: $\varepsilon_r = 72 \pm 1$ and $\sigma$ = 0.00557 and 0.00707 S/m. For HEPES solutions, the methodology provides $\varepsilon_r = 71$ and $\sigma$ = 0.00482 S/m. Again, to these significant digits, the same values of σ have been measured using the conductivity probe. We observed that for large concentrations of KCl the relative dielectric permittivity tends to decrease, this effect has been reported before. This has to do with the fact that $K^+$ and $Cl^-$ ions don't have an intrisec dipole moment (meaning a smaller polarizability than water), however they occupy space in the solution thus lowering the overall dielectric permittivity of the solution. Now we can say something about the measured frequency behavior of $Z_p$ with the increase of the KCl concentration (see Figure 4). The most notable things we observe are that the $ImZ_p$ has a power law behavior with an average exponent of 0.8 for uM concentrations, 0.6 for mM concentrations of KCl and 0.5 for HEPES. These are in good agreement with previous studies where, for mM concentrations of KCl, the average exponent was 0.7.[11] Moreover, the amplitude appears to depend on the concentration of KCl, i.e $ImZ_p$ goes as inverse proportional with the solution's conductivity, as expected.[11] Also the frequency where $Z_p$ becomes negligible increases with the ionic concentration. This last observation shows that polarization removal becomes more difficult at higher ionic concentrations.

c) *Colloidal suspensions.* We should point out from the beginning that the dielectric functions of colloidal suspensions are not constant with frequency. In fact, one of the main goals of dielectric spectroscopy is to capture and study these variations of ε with frequency. Thus, the methodology outlined above cannot be directly applied to these systems. However, one way to apply the methodology to colloids (including live cells) suspended in electrolytes or to samples saturated with electrolytes is as follows. Initially, the sample is measured and the raw values of $Z_{sample}$ are recorded. The data are, of course, contaminated by the polarization effect. Now the key observations are: 1) the polarization effect is due to the electrolyte and not the colloidal particles and 2) the dielectric function of the electrolyte is constant with frequency. Thus if we separate the electrolyte from the colloids, we can measure Zp as before, which then can be removed from $Z_{sample}$. Thus, *the following methodology emerges*:

α) Record the brute values of $Z_{sample}$.

β) Remove the colloids from the solution.

γ) Apply stepts i)-iv) to the supernatant and determine $Z_p$.

δ) Remove $Z_p$ from $Z_{sample}$ to obtain the intrinsic impedance of the sample.

In the case of live cell suspension, the step γ) can be done by a gentle centrifugation. Fast optical density (O.D.) measurements can be used to make sure there are no cells left in supernatant. The centrifugation should be gentle so that no cell rupture occurs during the process, otherwise the conductivity of the supernatant will be highly increased. An application of this methodology is shown in the right panel of Figure 3, to a suspension of live E-coli cells.

*In conclusion*, this paper presented a simple technique for the removal of polarization errors in dielectric measurements at low frequencies. In contrast to previous works, our methodology requires no assumption on the behavior of the polarization impedance, other than the widely accepted fact that the effect dies out at high frequencies. Critical to this method is that the frequency span of the measurements must incorporate a high frequency window where the polarization error becomes negligible. There is also no need for calibrations at any step in the experiment.

The method has been demonstrated to be robust for a class of electrolytes, including physiological buffers. Our study confirmed that the polarization impedance is reactive and that it varies as a power law with the frequency. Our study also showed that the amplitude and the exponent of the power law are weakly dependent on the applied voltage. However, $Z_p$ showed a dependence on the distance between the electrode plates, which becomes weaker for larger distances. For weak ionic solutions, we found that the exponent of the power law is independent while the amplitude decreases linearly with the increase of ionic concentration in line with previous work. Similar conclusions apply for strong ionic solutions. At last, the method was applied to live E.coli cell suspension. Obtaining uncontaminated dielectric spectroscopy data in alpha region opens the possibility of many interesting applications, the most notable being measuring the membrane potential as predicted in Ref.[9].

**Acknowledgement:** This work was supported by a grant from NJIT-ADVANCE which is funded by the National Science Foundation (grant # 0547427).

Figure Captions:

Figure 1. Real (1st row) and imaginary (2nd row) parts of the sample impedance for d=1mm (1st column), d=3mm (2nd column) and d=5mm (3rd column), for applied electric fields of 0.1 (red), 0.3 (blue), 0.5 (green), 0.7 (magenta) and 0.9 (cayan) V/cm. Black line marks $Z_{ideal}$.

Figure 2. Dielectric measurements for water and water with 1µM KCl, 5µM KCl, 10µM KCl, 20µM KCl: a) – e) real part of impedance; f) – j) imaginary part of impedance; k) – o) dielectric permittivity. Red lines marked with circles represent the dielectric values without compensating for the electrode polarization effects while blue, solid lines represent the dielectric values after removing the polarization effects. The green lines represent the imaginary part of the electrode polarization impedance.

Figure 3. Dielectric permittivity versus frequency for (left) water with 0.1 and 0.5 mM KCl and 5mM HEPES and (right) E.coli cell suspension for an optical density (O.D.) of 0.051. The dotted lines represent the real relative dielectric permittivity after the removal of the polarization error.

Figure 4. Imaginary part of the electrode polarization impedance for the samples from Figures 2 and 3 as function of frequency. The curves show a power law behavior with an average exponent equal to 0.8 for □M KCl concentrations, 0.6 for mM KCl concentrations and 0.5 for HEPES. The amplitude goes invers proportional with the conductivity, as discussed in text.

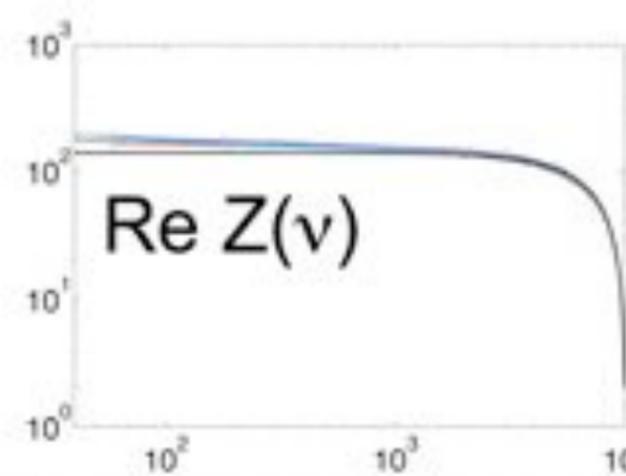 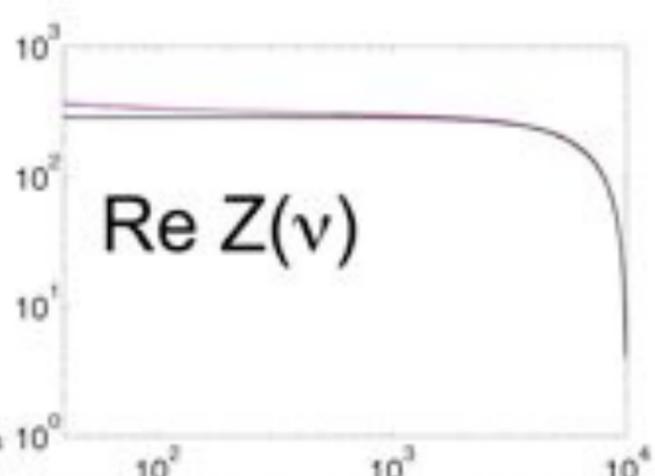 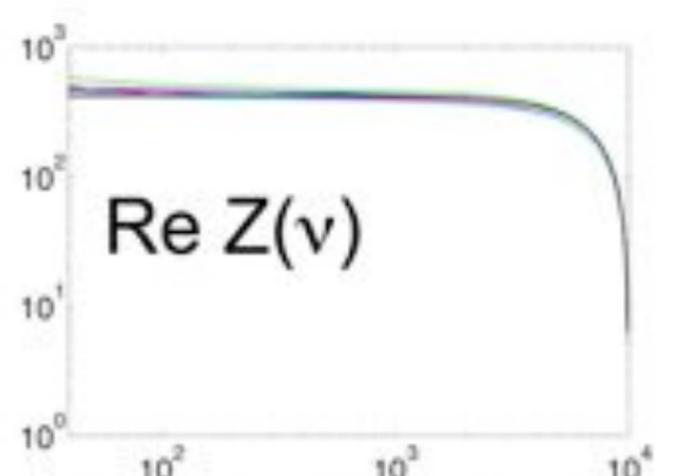
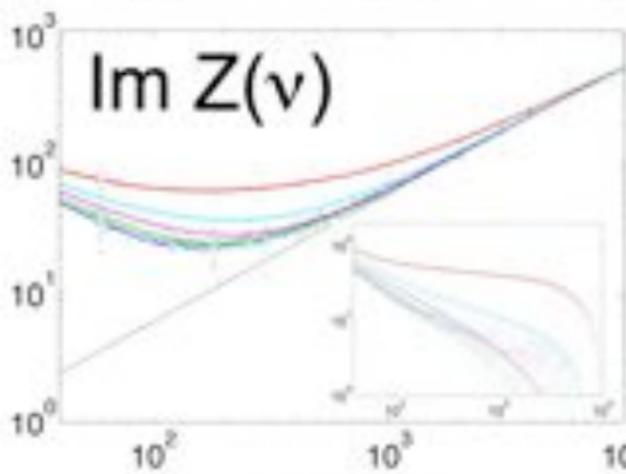 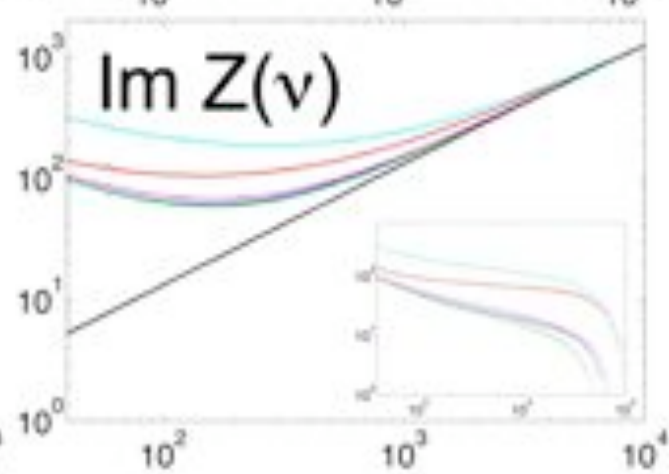 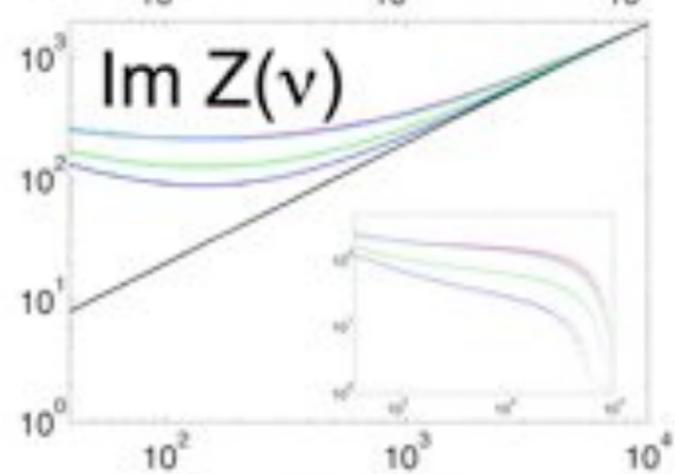

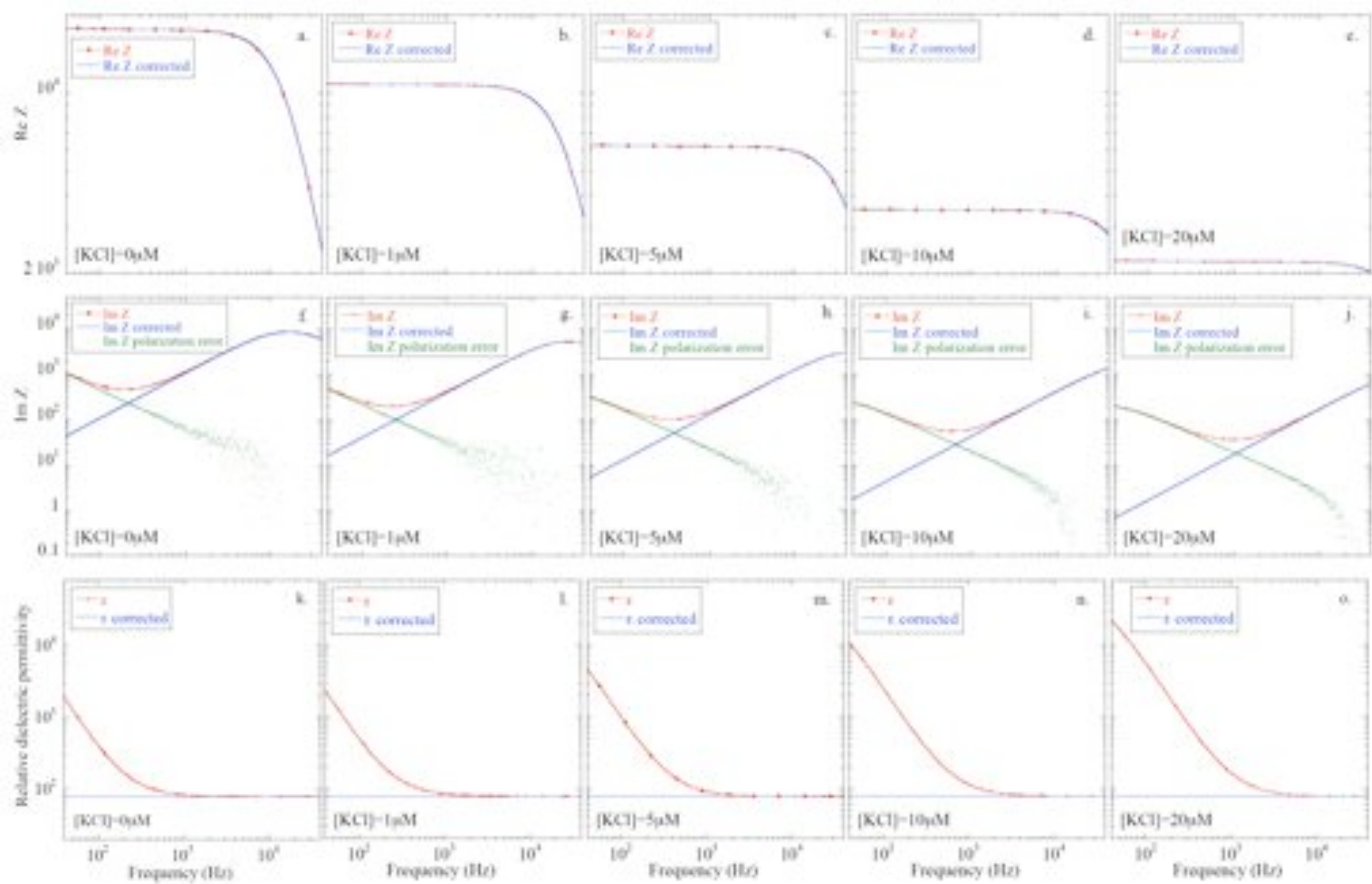

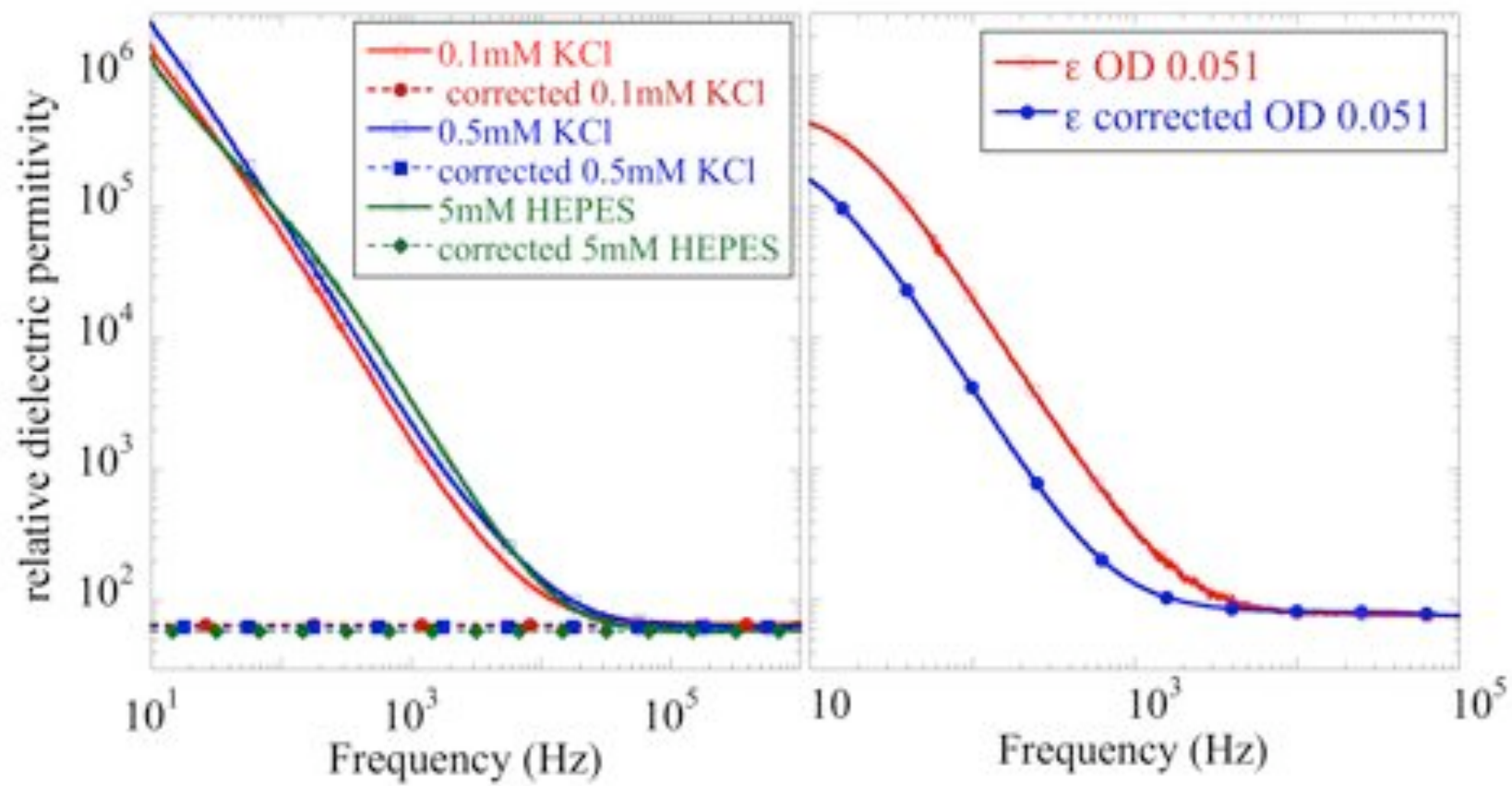

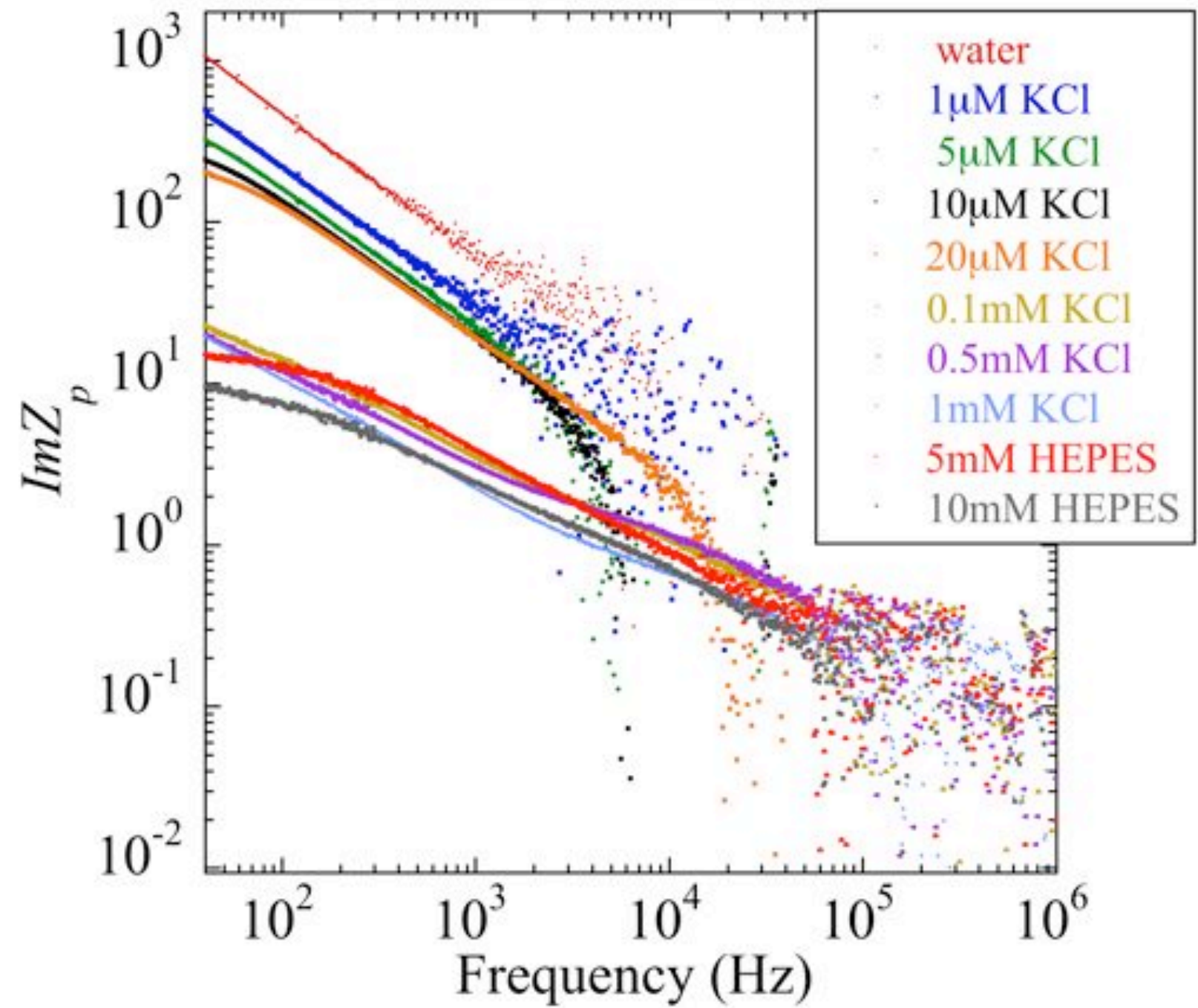